# Boron-doped diamond

Vassiliki Katsika-Tsigourakou[*]

*Section of Solid State Physics, Department of Physics, National and Kapodistrian University of Athens, Panepistimiopolis, 157 84 Zografos, Greece*

**Abstract-** Boron-doped diamond undergoes an insulator-metal transition at some critical value (around 2.21 at %) of the dopand concentration. Here, we report a simple method for the calculation of its bulk modulus, based on the thermodynamical model, by Varotsos and Alexopoulos, that has been originally suggested for the interconnection between the defect formation parameters in solids and bulk properties. The results obtained at the doping level of 2.6 at %, which was later improved at the level 0.5 at %, are in agreement with the experimental values.

*E-mail- vkatsik@phys.uoa.gr

## 1 INTRODUCTION

Pandey et al. [1] studied the pseudo elastic behavior of liquid alloys using pseudo potential model based on the density functional theory with both the local density approximation and the generalized gradient approximation for the exchange correlation function. Very interesting results were obtained which showed that the elastic constants of the elemental cubic model depend primarily on the bonding variance, the density at the cell boundary and the symmetry of the lattice.

In the above paper Pandey et al. applied the model of Varotsos and Alexopoulos using slight modification in volume due to concentration. It is the scope of this short paper to extend the usefulness of the challenging findings of Pandey et al. and show that the use of Varotsos and Alexopoulos model can also serve for treating a problem (see below) of major technological interest.

Diamond has been extensively studied (e.g., see refs [2-12]) in view of its remarkable properties. For example, it has a very large Debye temperature [13] and the largest elastic moduli known for any material and correspondingly the largest sound velocities [13], [14]. Nowdays, the diamond anvil cell (DAC) technique, which is extensively used as a unique tool for producing high pressure in the laboratory [15], exploits the extreme hardness of diamond.

Diamond is a wide band-gap semiconductor. The high interest of studying both doped natural diamonds and high-level doped synthetic diamonds [16] originates from the discovery of the profound influence of dopants on their physical properties. Specifically, doping diamond with boron leads to the insulator-metal transition [17]. Electrical conductivity measurements of diamond revealed that for boron concentrations higher than some critical value estimated as 2.21 at %, the conductivity on the metallic side of the transition at low temperature a $T^b$ law. For metallic samples, b was found to be 1/3, approaching 1/2 at higher concentrations [17]. Some uncertainty remains in predicting the boron concentration above which metallic conduction takes place [17-21].

The isothermal bulk modulus B (and the compressibility $\kappa, \kappa = 1/B$) can be used as a quantitative characteristic describing relations between the structure and atomic forces, from one side and the physical properties of solids, from another side. Dubrovinskaia et al. [22] reported the results of high pressure-high temperature synthesis of boron-doped diamond and the results of experimental determination of its bulk modulus. In addition, they proceeded to detailed a theoretical calculation which suggested very little (within computational uncertainty) effect of the doping on the compressibility of diamond for impurity concentrations up to 3 at %. These calculations also confirmed that boron atoms prefer to substitute C-atoms in a diamond structure. It is the aim of this short paper to draw attention to the following point: Instead of the aforementioned tedious theoretical calculation, a simple thermodynamical model can be alternatively used for the estimation of the boron-concentration dependence of the compressibility of diamond. This thermodynamical model, has been originally suggested for the formation and/or migration processes of defects in solids [23-25]. The same model was extended [26] to describe the physical properties of the electric signals that precede rupture [27-29]. In the next section, we recapitulate this model (termed $cB\Omega$ model, see below) and then in section 3 we apply it to the case of the compressibility of boron-doped diamond. We note that the success of this model to reproduce the self-diffusion coefficients of diamond has been already checked in [30].

## 2 THE MODEL

Let us denote $V_1$ and $V_2$ the corresponding molar volumes, i.e. $V_1 = N\upsilon_1$ and $V_2 = N\upsilon_2$ (where $N$ stands for Avogadro's number) for diamond (density 3.51

gr/cm$^3$) and B$_4$C (density 2.48 gr/cm$^3$) respectively. Defining a "defect volume" [31] $\upsilon^d$ as the variation of the volume $V_1$, if one "molecule" of type "1" is replaced by one "molecule" of type "2", it is evident that the addition of one "molecule" of type "2" to a crystal containing $N$ "molecules" of type "1" will increase its volume by $\upsilon^d + \upsilon_1$. Assuming that $\upsilon^d$ is independent of composition, the volume $V_{N+n}$ of a crystal containing $N$ "molecules" of type (1) and $n$ "molecules" of type "2" can be written as [27,31]:

$$V_{N+n} = [1+(n/N)]V_1 + n\upsilon^d \qquad (1)$$

The compressibility $\kappa$ of the doped diamond can be found by differentiating (1) with respect to pressure which finally gives [27]:

$$\kappa V_{N+n} = \kappa_1 V_1 + (n/N)\left[\kappa^d N \upsilon^d + \kappa_1 V_1\right] \qquad (2)$$

where $\kappa^d$ denotes the compressibility of the volume $\upsilon^d$, defined as [32] $\kappa^d \equiv -(1/\upsilon^d) \times (d\upsilon^d/dP)_T$.

Within the approximation of the hard-spheres model, the "defect-volume" $\upsilon^d$ can be estimated from:

$$\upsilon^d = (V_2 - V_1)/N \qquad (3)$$

Thus, if $V_{N+n}$ can be determined from (1) (upon either considering (3) or other type of measurements and/or method), the compressibility $\kappa$ can be found from (2) if a procedure for the estimation of $\kappa^d$ will be employed. In this direction, we adopt a thermodynamical model, termed $cB\Omega$ model, for the formation and migration of the defects in solids [23-27]. According to this model, the defect Gibbs energy $g^i$ is interconnected with the bulk properties of the solid through the relation $g^i = c^i B \Omega$ where $B$ stands, as mentioned, for the isothermal bulk modulus ($=1/\kappa$), $\Omega$ the mean volume per atom and $c^i$ is dimensionless quantity. (The superscript $i$ refers to the defect process under consideration, e.g. defect formation, defect migration and self-diffusion activation). By differentiating this relation in respect to pressure $P$, we find the defect volume $\upsilon^i$ $[=(dg^i/dP)_T]$. The compressibility $\kappa^{d,i}$ defined by $\kappa^{d,i}$ $[\equiv -(d\ell n\upsilon^i/dP)_T]$, is given by:

$$\kappa^{d,i} = (1/B) - (d^2B/dP^2)/[(dB/dP)_T - 1] \qquad (4)$$

This relation states that the compressibility $\kappa^{d,i}$ does *not* depend on the type i of the defect process. Thus, it seems reasonable to assume [32], [27] that the validity of (4) holds also for the compressibility $\kappa^d$ involved in (2), i.e.,

$$\kappa^d = \kappa_1 - (d^2B_1/dP^2)/[(dB_1/dP)_T - 1] \qquad (5)$$

where the subscript <1> in the quantities at the right side denotes that they refer to the undoped diamond crystal.

## 3 APPLICATION OF THE MODEL TO THE BORON-DOPED DIAMOND

In general, the quantities $dB_1/dP$ and $d^2B_1/dP^2$, can be roughly estimated from the modified Born model according to [27], [31]:

$$dB_1/dP = (n^B + 7)/3 \quad \text{and} \quad B_1(d^2B_1/dP^2) = -(4/9)(n^B + 3) \qquad (6)$$

where $n^B$ is the usual Born exponent. In cases, however, where the Born model does not provide an adequate description, we can solely rely on (4), but not on (6). In other words, if Born model holds, we calculate the first and second pressure derivatives of the bulk modulus on the basis of (6) and then insert them into (4). Otherwise, we insert into (4) the first and second pressure derivative of the bulk modulus deduced from the elastic data under pressure (obtained either from laboratory measurements or from microscopic calculations) using a least squares fit to a second order Murnaghan equation. The results of these possibilities are now described below for the boron-doped diamond. We shall use hereafter the experimental value $B_1$=442 GPa obtained in Refs [8], [22], [33] for the pure diamond crystal.

In Dubrovinskaia et al. [22], the experimental pressure-volume data were fitted using the third order Birch- Murnaghan equation of state. This fitting procedure for (undopted) diamond gave $B_1$=442(4) GPa, $B_1'$=3.2(2) –where $B_1'$ denotes the first pressure derivative of $B_1$- and the zero- pressure volume 3.4157(9) cm$^3$/mol which within the uncertainty coincide with the data from [8]. For boron-doped diamond, at the doping level of 2.6 at % they found $B$=436(7) GPa, $B'$=3.1(2) and the zero-pressure volume 3.4319(9) cm$^3$/mol.

Using the aforementioned values for the zero- pressure volume of the undoped and the boron-doped diamond we determined $\upsilon^d$ on the basis of (1). Let us now investigate the $\kappa(=1/B)$ value resulting from (2) at the doping level of 2.6 at % when employing the determination of the $\kappa^d$ -value by means of the procedures described above:

First, when employing the modified Born model –and hence use (6)- we find $B \approx 433$ GPa. This is marked with open circle in Fig. 1 and is very close to the experimental value $B$=436(7) GPa measured in [22] (marked with cross in Fig. 1).

Second, we now employ the Morse potential parameters determined –in the frame of an analytic mean field approach- by fitting the compression experimental data of diamond at ambient temperature. For example see Fig. 3 of [34] which, if described by an expansion of the isothermal bulk modulus carried out to second order, gives $B_1'=3.3$ and $B_1''=-0.0024$ GPa$^{-1}$. By inserting these $B_1'$ and $B_1''$ values into (5) we find $\kappa^d$ and then from (2) we get $B=433$ GPa (marked with open square in Fig. 1). This is also very close to the experimental value $B=436(7)$ GPa reported in [22]. Finally, we note that the latter calculation was repeated by using, instead of the aforementioned doping level 2.6 at %, the value of 0.5 at % that was later reported [35] as being more representative of the reality since it was deduced after closer investigations of the microstructure of boron-doped diamond and of boron distribution. This led to the calculated value $B=441.9$ GPa, marked with an inverted triangle in Fig. 1, which also agrees with the experimental results, if the experimental error is considered.

## 4 CONCLUSION

In summary, for boron-doped diamond, at doping level of 2.6 %, Dubrovinskaia et al [22] reported the experimental value of the isothermal bulk modulus $B=436$ GPa. The values of $B(=1/\kappa)$ calculated here on the basis of (2) are found to be 433 GPa when using the $\kappa^d$-value obtained from (4) in terms of $B_1'$ and $B_1''$, of (undoped) diamond estimated from the either modified Born model or its analytic equation of state based on an analytic mean field approach. In view of a large error in the boron-content, the position of this experimental point in Fig. 1 can be found [33] at a concentration as low as 2 at %. In this case the calculated $B$-values from (2) are found to be around 435 GPa, thus being again in very good agreement with the experimental value [22] of 436 GPa. If we alternatively use an even lower concentration of 0.5 at %, which was later [35] reported as being closer to the reality, our calculated value is around 441.9 GPa which also agrees with the experimental results, if the experimental uncertainty is considered. For the sake of comparison, we note that the calculated $B$-value in the framework of the density functional method [33] is around 421 GPa.

**Figure & Figure Caption**

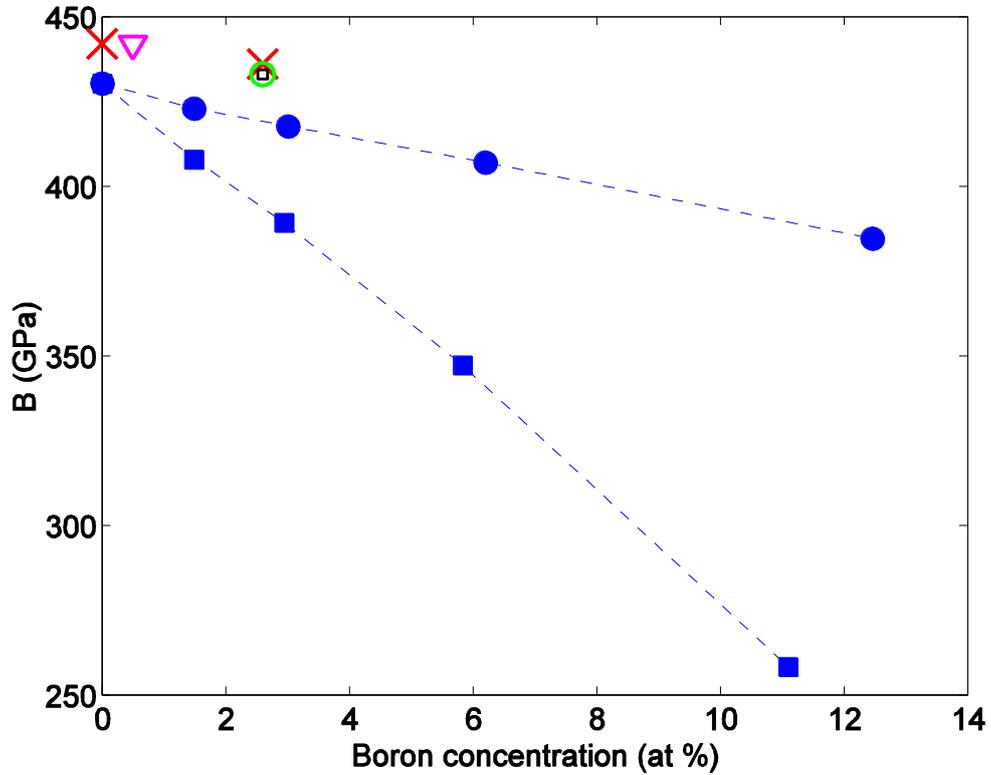

**Fig. 1.** Dependence of the isothermal bulk modulus on the boron concentration. Experimental values obtained in [22], [33] are shown with crosses. Calculated results of the bulk moduli with substitutional (solid circles) and interstitials (squares) boron impurities according to [22], [33]. The results calculated in this paper are designated with open circles and open squares when employing (2) and using the compressibility $\kappa^d$ of the defect volume obtained either from the modified Born model or from the analytic equation of state in [34] described in the text. The latter calculation was repeated by considering, instead of 2.6 at %, the more recent value 0.5 at % reported in [35], and led to the value $B=441.9$ GPa shown by the inverted triangle.